\documentclass[a4paper,11pt]{article}
\usepackage[utf8]{inputenc}
\usepackage{pos}
\usepackage{amsmath}
\usepackage{axodraw2}

\title{Short-distance constraints in hadronic-light-by-light for the muon $g-2$}

\author*[a]{Johan Bijnens}
\author[b]{Nils Hermansson-Truedsson}
\author[c]{Antonio Rodríguez Sánchez}
\affiliation[a]{Department of Astronomy and Theoretical Physics,
Lund University,\\ 
Box 43, SE221-00 Lund, Sweden}

\affiliation[b]{Albert Einstein Center for Fundamental Physics, Institute for Theoretical Physics,Universit\"{a}t Bern,\\
  Sidlerstrasse 5, CH–3012 Bern, Switzerland}

\affiliation[c]{Universit\'{e} Paris-Saclay, CNRS/IN2P3, IJCLab,\\91405 Orsay,
France}

\emailAdd{johan.bijnens@thep.lu.se}
\emailAdd{nils@itp.unibe.ch}
\emailAdd{arodriguez@ijclab.in2p3.fr}

\abstract{%
In this talk recent progress in studying the short-distance properties of the hadronic light-by-light contribution to the muon $g-2$ is described. The intermediate and short-distance part is a major contributor to the error of the theoretical prediction as described in the recent white paper \cite{Aoyama:2020ynm}. We have shown that the massless quark-loop is the first term in a systematic expansion at short-distances, a result already used in the white paper. Newer results conclude that both nonperturbative and perturbative corrections are under control. The talk describes these developments and how they fit in the total theoretical prediction for the muon $g-2$.}

\renewcommand{\logo}
{\raisebox{-5mm}
{\parbox{\textwidth}{\begin{flushright}
LU TP 21-47\\
October 2021\\
\end{flushright}
\includegraphics{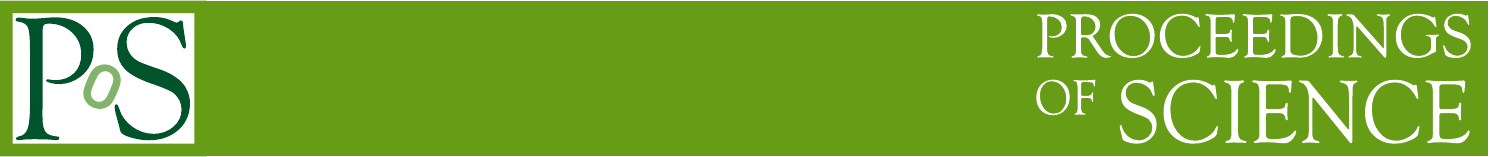}}
}}

\FullConference{%
  *** Particles and Nuclei International Conference - PANIC2021 ***\\
  *** 5 - 10 September, 2021 ***\\
  *** Online ***
}

\newlength{\bibitemsep}
\newlength{\bibparskip}
\let\oldthebibliography\thebibliography
\renewcommand\thebibliography[1]{%
  \oldthebibliography{#1}%
  \setlength{\parskip}{\bibitemsep}%
  \setlength{\itemsep}{\bibparskip}%
}
\makeatletter
\renewcommand\section{\@startsection{section}{1}{\z@}%
                                   {2.0ex \@plus 0.ex \@minus 0.35ex}%
                                   {1.0ex \@plus 0.ex \@minus  0.2ex}%
                                   {\normalfont\large\secstyle}}
\makeatother

\begin{document}
\maketitle

\section{Introduction}

There is a long standing difference between the Standard Model (SM) prediction~\cite{Aoyama:2020ynm} for the muon anomalous magnetic moment $a_\mu=(g-2)/2$ and the experimental measurement~\cite{Bennett:2006fi} at BNL. The latter was recently confirmed by the experiment at FNAL~\cite{Muong-2:2021ojo,Fertlpanic}. The theoretical
prediction, $a_\mu^{SM}$, and experimental average, $a_\mu^{exp}$, give a difference $\Delta a_\mu$:
\begin{align}
  a_\mu^{SM}&= 116 591 810(43)\times10^{-11}, &
  a_\mu^{exp}&= 116 592 061(41)\times10^{-11}, &
\Delta a_\mu&= 251(59)\times10^{-11}.
\end{align}
The difference is 4.2$\sigma$. The precision at FNAL is expected to improve significantly and there is an independent measurement planned at J-PARC. Improving the theoretical prediction is needed.

\begin{figure}
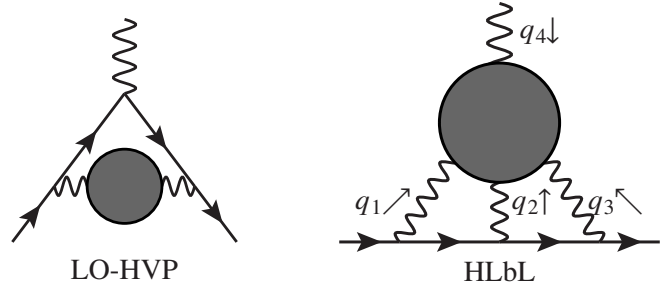

\begin{center}  
\setlength{\unitlength}{1.4pt}
\begin{axopicture}(60,70)(0,-10)
\SetScale{1.4}
\SetWidth{0.75}
\Photon(30,60)(30,40){3}{3.5}
\ArrowLine(0,0)(11.25,15)
\ArrowLine(11.25,15)(30,40)
\ArrowLine(30,40)(48.75,15)
\ArrowLine(48.75,15)(60,0)
\Photon(11.25,15)(20,15){-2.5}{2.}
\Photon(40,15)(48.75,15){2.5}{2.}
\GCirc(30,15){10}{0.4}
\Text(30,-10)[b]{LO-HVP}
\end{axopicture}
\hspace*{2cm}
\setlength{\unitlength}{1.5pt}
\begin{axopicture}(90,80)(0,-10)
\SetScale{1.5}
\SetWidth{0.75}
\Photon(40,60)(40,40){3}{3.5}
\ArrowLine(0,0)(15,0)
\ArrowLine(15,0)(40,0)
\ArrowLine(40,0)(65,0)
\ArrowLine(65,0)(80,0)
\Photon(15,0)(40,40){2}{9}
\Photon(40,0)(40,40){-2}{8}
\Photon(65,0)(40,40){-2}{9}
\GCirc(40,30){15}{0.4}
\Text(18,10)[r]{$q_1\!\!\nearrow$}
\Text(43,10)[l]{$q_2\!\!\uparrow$}
\Text(62,10)[l]{$q_3\!\!\nwarrow$}
\Text(45,53)[l]{$q_4\!\!\downarrow$}
\Text(40,-10)[b]{HLbL}
\end{axopicture}
\end{center}
\caption{The LO-HVP and HLbL contributions shown schematically. The arrowed line is the muon, wiggly lines indicate photons and the blobs indicate hadronic contributions.}
\label{fig:hadronic}
\end{figure}

The theoretical error is dominated by two hadronic contributions, the lowest order hadronic vacuum polarization (LO-HVP) and the hadronic light-by-light contribution (HLbL), shown schematically in Fig.~\ref{fig:hadronic}.
The higher order hadronic, electroweak and QED contributions are sufficiently precise to not significantly contribute to the theoretical error. The error at the moment is dominated by the LO-HVP but here improvements on both the dispersive and lattice calculations are expected. We will not discuss further the hadronic vacuum polarization determinations.

The HLbL contribution and its error as estimated in the white paper \cite{Aoyama:2020ynm} are
\begin{align}
  \label{eq:HLbLWP}
a_\mu^{HLbL} = 92(18)~10^{-11}\,.
\end{align}
There have been a number of improvements since then, including an improved lattice QCD calculation \cite{Chao:2021tvp} compatible both with the earlier lattice results \cite{Blum:2019ugy} and (\ref{eq:HLbLWP}). The main problem with the HLbL calculation is that, as shown in the right figure of Fig.~\ref{fig:hadronic}, its evaluation involves always one very low momentum, $q_4$ corresponding to the external magnetic field and momenta $q_1,q_2,q_3$ which span the entire range, both low and high. For a long time this only allowed for model calculations, see e.g.
\cite{Bijnens:1995cc,Bijnens:1995xf,Hayakawa:1997rq,Bijnens:2001cq,Hayakawa:2001bb,Prades:2009tw}. More recently \cite{Colangelo:2015ama,Colangelo:2017fiz} produced a dispersive method allowing the long-distance parts to get under better control. Tab.~\ref{tab:HLbLmain} summarizes the phenomenological parts of the HLbL as put together in the white paper~\cite{Aoyama:2020ynm}.
\begin{table}
  \begin{tabular}{lrlr}
    \hline
    \multicolumn{2}{c}{Long distance}
    &\multicolumn{2}{c}{Short and medium distance}\\
    \hline
    $\pi^0$ (and $\eta,\eta^\prime$) pole & $ 93.8(4.0)~10^{-11}$ &
    Charm (beauty, top) loop & $ 3(1)~10^{-11}$\\
 Pion and kaon box (pure) & $ -16.4(2)~10^{-11}$ &
 Axial vector &  $ 6(6)~10^{-11}$\\
 $\pi\pi$-rescattering &$ -8(1)~10^{-11}$ &
                                                                              Short-distance & $ 15(10)~10^{-11}$\\
    \hline
  \end{tabular}
  \caption{The main contributions to the phenomenological evaluation of HLbL as estimated in~\cite{Aoyama:2020ynm}. Scalars below 1~GeV are included in $\pi\pi$-rescattering. Scalars above 1~GeV are small.}
  \label{tab:HLbLmain}
\end{table}
As one can see the long distance and heavier quark contributions are under good control. The axial-vectors and the short-distance part provide the bulk of the error. These were added linearly in \cite{Aoyama:2020ynm} and include a guestimate of the overlap between the short-distance from the quark-loop and the other contributions as well as of the contribution from other resonances above 1~GeV. The work we describe here~\cite{Bijnens:2019ghy,Bijnens:2020xnl,Bijnens:2021jqo} should allow to reduce this error.

Short-distance constraints can be used in many ways, here we concentrate on those relevant for the entire four-point function (\ref{eq:fourpoint}) and its derivative at $q_4=0$.

The underlying object in HLbL is the hadronic blob in Fig.~\ref{fig:hadronic} with four photons attached to it, i.e. the four-point function of four electromagnetic currents:
\begin{align}
  \label{eq:fourpoint}
\Pi^{\mu\nu\lambda\sigma}= -i\int\!d^4x d^4y d^4z
e^{-i(q_1\cdot x+q_2\cdot y+q_3\cdot z)}
\left\langle T\left(j^\mu(x) j^\nu(y) j^\lambda(z) j^\sigma(0)\right)\right\rangle
\end{align}
In the notation of \cite{Colangelo:2017fiz} the contribution to $a_\mu^{\text{HLbL}}$ comes via 
\begin{align}
  \Pi^{\mu\nu\lambda\sigma}&= \sum_{i=1}^{54} T_i^{\mu\nu\lambda\sigma}\hat\Pi_i,&
  \left. {\delta \Pi^{\mu\nu\lambda\sigma}\over \delta  q_{4\rho}}\right|_{q_4=0}&=  \sum_{i=1}^{54}\left. {\delta T_i^{\mu\nu\lambda\sigma}\over \delta  q_{4\rho}}\hat\Pi_i\right|_{q_4=0},&  Q_3^2 &= Q_1^2+Q_2^2+2Q_1 Q_2\tau,\\
 a_\mu^{\text{HLbL}} &= 
  \frac{2\alpha^3}{3\pi^2}\int_0^\infty\!\!\! dQ_1 dQ_2
Q_1^3 Q_2^3
  \int_{-1}^1\!\!\!d\tau
         \sqrt{1-\tau^2}\sum_{i=1}^{12}\hat T_i\left(Q_1,Q_2,\tau\right)\overline \Pi_i\left(Q_1,Q_2,\tau\right).\hspace*{-9cm}
\end{align}
The 12 $\overline\Pi_i$ can be obtained from the $\hat\Pi_i$ for $i=1,4,7,17,39,54$. We use $Q_i\cdot Q_j=-q_i\cdot q_j$.

\section{Constituent quark-loop}

The constituent quark-loop has been used for full HLbL estimates since the 1970s and has often been recalculated. It was used in \cite{Bijnens:1995xf} to match with short-distances using the mass as a lower-cutoff.
\begin{figure}
\begin{center}
  \begin{minipage}{80pt}
\begin{axopicture}(70,70)
\Photon(0,0)(20,20){2.5}{3.5}
\Photon(0,70)(20,50){2.5}{3.5}
\Photon(70,70)(50,50){2.5}{3.5}
\Photon(70,0)(50,20){2.5}{3.5}
\ArrowLine(20,20)(50,20)
\ArrowLine(50,20)(50,50)
\ArrowLine(50,50)(20,50)
\ArrowLine(20,50)(20,20)
\Text(60,15)[l]{$q_4$}
\Text(10,15)[r]{$q_1$}
\Text(10,55)[r]{$q_2$}
\Text(60,55)[l]{$q_3$}
\Arc[arrow](35,35)(9,-80,260)
\Text(35,35){$p$}
\end{axopicture}\\
\centerline{(a)}
\end{minipage}
~~~
\begin{minipage}{70pt}
\begin{axopicture}(70,70)
\Photon(0,0)(20,20){2.5}{3.5}
\Photon(0,70)(20,50){2.5}{3.5}
\Photon(70,70)(50,50){2.5}{3.5}
\Photon(70,0)(50,20){2.5}{3.5}
\ArrowLine(50,20)(50,33)
\ArrowLine(50,37)(50,50)
\Vertex(50,33){1}
\Vertex(50,37){1}
\ArrowLine(50,50)(20,50)
\ArrowLine(20,50)(20,20)
\Text(60,15)[l]{$q_4$}
\Text(10,15)[r]{$q_1$}
\Text(10,55)[r]{$q_2$}
\Text(60,55)[l]{$q_3$}
\SetWidth{2.0}
\ArrowLine(20,20)(50,20)
\end{axopicture}\\
\centerline{(b)}
\end{minipage}
~~~~~~
\begin{minipage}{70pt}
\begin{axopicture}(60,70)
\Photon(0,0)(15,10){2.5}{3.5}
\ArrowLine(15,10)(35,10)
\ArrowLine(35,10)(55,10)
\Text(35,10){$\otimes$}
\Text(35,5)[t]{``$q_4$''}
\ArrowLine(55,10)(35,40)
\ArrowLine(35,40)(15,10)
\Photon(55,10)(70,0){2.5}{3.5}
\Photon(35,40)(35,55){2.5}{3.5}
\Text(5,10)[r]{$q_1$}
\Text(65,10)[l]{$q_2$}
\Text(38,48)[l]{$q_3$}
\Arc[arrow](35,22)(7,-80,260)
\Text(35,21){$p$}
\end{axopicture}\\
\centerline{(c)}
\end{minipage}
~~~~~~
\begin{minipage}{70pt}
\begin{axopicture}(60,70)
\Photon(0,0)(15,10){2.5}{3.5}
\ArrowLine(15,10)(35,10)
\ArrowLine(35,10)(55,10)
\Text(35,10){$\otimes$}
\Text(35,5)[t]{``$q_4$''}
\ArrowLine(55,10)(45,25)
\ArrowLine(45,25)(35,40)
\ArrowLine(35,40)(25,25)
\ArrowLine(25,25)(15,10)
\Gluon(25,25)(45,25){-2.5}{4}
\Photon(55,10)(70,0){2.5}{3.5}
\Photon(35,40)(35,55){2.5}{3.5}
\Text(5,10)[r]{$q_1$}
\Text(65,10)[l]{$q_2$}
\Text(38,48)[l]{$q_3$}
\end{axopicture}\\
\centerline{(d)}\end{minipage}
\end{center}
\caption{(a) The (constituent) quark-loop. (b) The quark-loop with an insertion of the quark-antiquark vacuum-expectation-value. (c) An insertion of the back-ground field. (d) An example of a gluonic correction.}
\label{fig:quark-loop}
\end{figure}
\begin{figure}
\begin{minipage}{6cm}
  \includegraphics[width=0.99\textwidth]{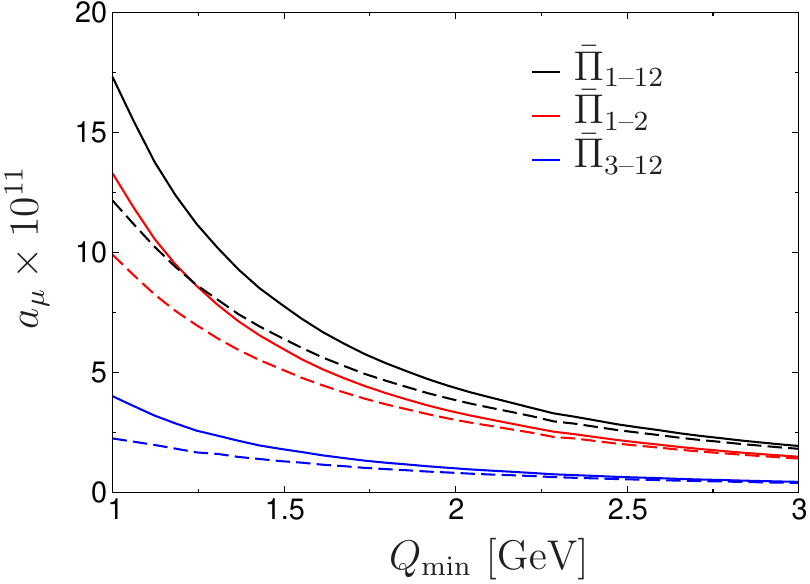}\\
\centerline{(a)}
\end{minipage}
\begin{minipage}{8cm}
  \includegraphics[width=0.99\textwidth]{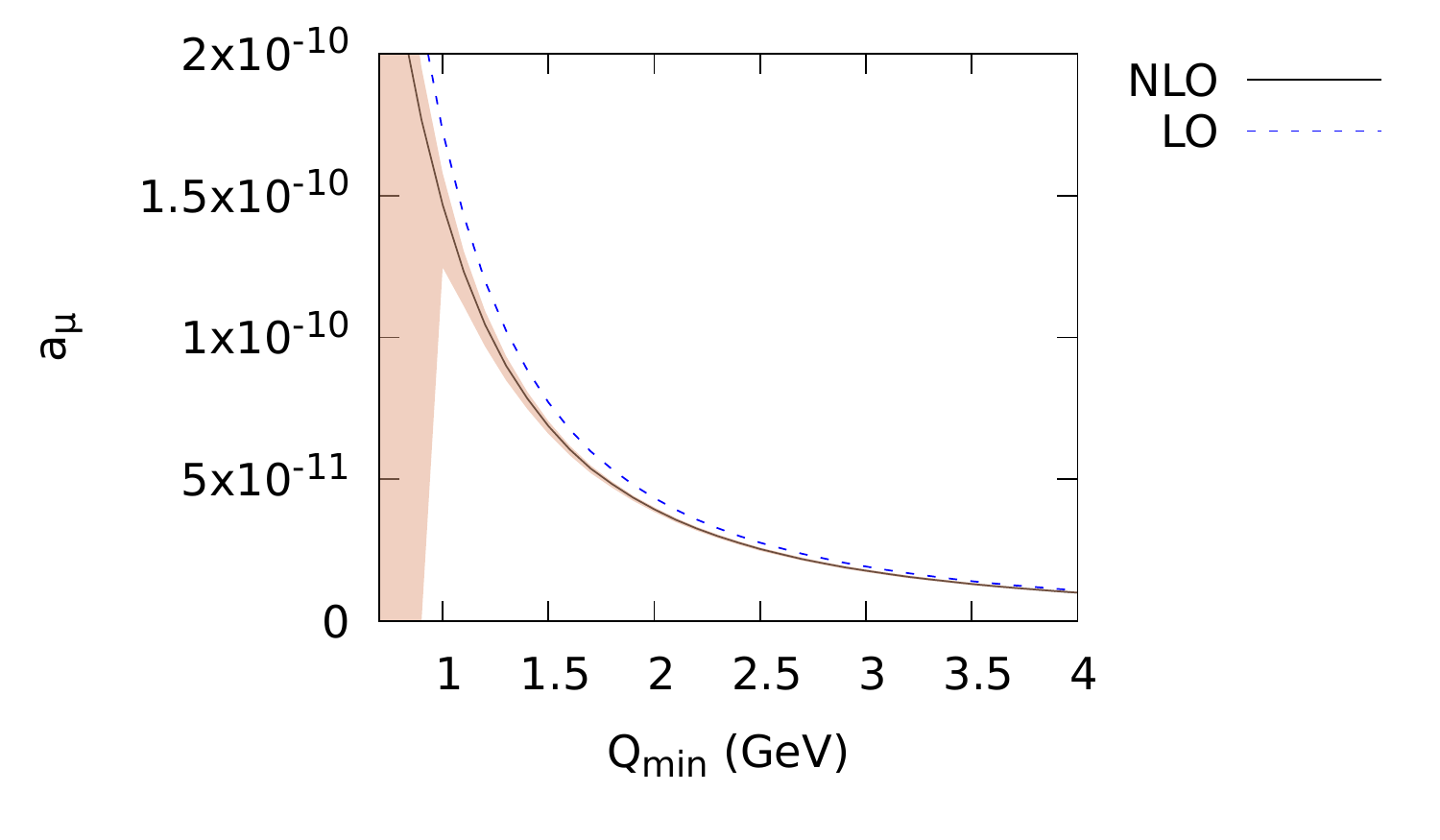}\\[-4mm]
\centerline{(b)}
\end{minipage}\\[-3mm]
\caption{The contribution with $Q_i\ge Q_{\text{min}}$ for (a) $M_Q=0$ and $M_Q=0.3$~GeV, from \cite{Aoyama:2020ynm}; (b)
the massless quark-loop (LO) and the gluonic correction (NLO) and its uncertainty due to varying $\alpha_S$, from \cite{Bijnens:2021jqo}.}
\label{fig:quark-loopnum}
\end{figure}

The total contribution with $M_Q=0.3$~GeV is $a_\mu^{\text{HLbLQ}}=54\cdot10^{-11}$ and above 1 GeV only $12\cdot10^{-11}$. The massless quark-loop above 1~GeV is $17\cdot10^{-11}$. One of our results is that the latter makes sense within QCD as the first term in a well defined operator-product-expansion (OPE). The dependence on $Q_{\text{min}}$ is shown in Fig.~\ref{fig:quark-loopnum}(a) and goes as $1/Q_{\text{min}}^2$ as expected.

\section{First attempt: naive OPE}

The usual OPE applied to (\ref{eq:fourpoint}) puts all currents close together or assumes that all $|Q_i\cdot Q_j|$ are large. We then take the derivative w.r.t. $q_4$ and send $q_4\to 0$.  The simple quark-loop of Fig.~\ref{fig:quark-loop}(a) is well defined, with $Q_1^2,Q_2^2,Q_3^2\ge Q_{\text{min}}^2\gg \Lambda_{\textrm{QCD}}^2$. The loop integration over $p$ damps the infrared (IR) divergence. The problem comes when we try to add higher orders in the OPE, e.g. a vacuum expection value contribution shown in Fig.~\ref{fig:quark-loop}(b). The thick propagator becomes divergent when we send $q_4\to0$. This method cannot be used to obtain a proper OPE for the HLbL contribution to $a_\mu$.

\section{OPE in an background field}

The same problem with limits appeared in the QCD sum rule calculations for electromagnetic radii and magnetic moments \cite{Balitsky:1983xk,Ioffe:1983ju}. There the solution was to do the OPE in a background field. We thus treat the $q_4$-leg as a constant background field and do the OPE in its presence. The IR divergences are absorbed in condensates and the expansion where the three remaining currents are close or $Q_1^2,Q_2^2,Q_3^2\gg\Lambda_{\text{QCD}}^2$ can be done. An example of a background induced condensate is the magnetic quark-susceptibility $X_q$ defined via
$\langle \bar q\sigma_{\alpha\beta} q\rangle \equiv  e_q F_{\alpha\beta} X_q\,$.  
Pictorially one replaces the diagram of Fig.~\ref{fig:quark-loop}(b) with the one of  Fig.~\ref{fig:quark-loop}(c) and the IR divergence of the zero-momentum propagator in Fig.~\ref{fig:quark-loop}(b) is absorbed in the induced condensates.

This expansion can be done in principle to any order. The first term is the massless quark-loop~\cite{Bijnens:2019ghy}, the next term is proportional to the quark-mass, due to helicity, and the magnetic susceptibility. More details, including the detailed treatment of the IR divergences, and the expansion including the first three orders is done in \cite{Bijnens:2020xnl}. As an example of the IR interplay, the $m_q^2$ corrections to the quark-loop mix with the magnetic susceptibility contribution.

Using phenomenological estimates or lattice calculations, values for all condensates were obtained in \cite{Bijnens:2020xnl}. The contribution to $a_\mu$ is shown in Tab.~\ref{tab:OPE}. The main message is that all higher orders are small because of small quark-masses and condensates for a lower cut-off above 1~GeV.
\begin{table}
\begin{minipage}{9cm}
  \begin{tabular}{|l|l|r|r|}
\hline
  Order &
   Contribution 
   & $Q_{\textrm{min}}=$ 
   & $Q_{\textrm{min}}=$ 
   \\
   &
   
   & $1 \,\mathrm{GeV}$ 
   & $2 \,\mathrm{GeV}$ 
   \\ \hline
\rule{0pt}{10pt}$1/Q_{\textrm{min}}^2$ &   quark-loop
   & $1.73\cdot 10^{-10}$               
   & $4.35\cdot 10^{-11}$                
   \\ \hline
    \rule{0pt}{11pt}$1/Q_{\textrm{min}}^4$ & \begin{minipage}{2cm}quark-loop,\\\rule{0.5cm}{0cm}~$m_q^2$\end{minipage}
   & $-5.7 \cdot 10^{-14}$               
   & $-3.6 \cdot 10^{-15}$                
   \\
 &  $X_{2,m}$    
   & $-1.2\cdot 10^{-12}$               
   & $-7.3\cdot 10^{-14}$                
   \\ \hline
\rule{0pt}{11pt} $1/Q_{\textrm{min}}^6$ &  $X_{2,m^3}$    
   & $6.4\cdot 10^{-15}$
   & $1.0\cdot 10^{-16}$                
   \\
&   $X_{3}$      
   & $-3.0\cdot 10^{-14}$
   & $-4.7\cdot 10^{-16}$             
   \\
 &  $X_{4}$      
   & $3.3\cdot 10^{-14}$ 
   & $5.3\cdot 10^{-16}$
   \\
&   $X_{5}$      
   & $-1.8\cdot 10^{-13}$ 
   & $-2.8\cdot 10^{-15}$ 
   \\
 &  $X_{6}$      
   & $1.3 \cdot 10^{-13}$    
   & $2.0\cdot 10^{-15}$         
   \\
 &  $X_{7}$      
   & $9.2\cdot 10^{-13}$     
   & $1.5\cdot 10^{-14}$   
   \\
 &  $X_{8,1}$    
  & $3.0\cdot 10^{-13}$             &    $4.7\cdot 10^{-15}$         \\
& $X_{8,2}$ 
 &   $-1.3\cdot 10^{-13}$              &  $-2.0\cdot 10^{-15}$ \\  
\hline
  \end{tabular}\\
  \centerline{(a)}
\end{minipage}
\begin{minipage}{6cm}
\begin{tabular}{|l|r|r|}
  \hline        & Quark- & Gluon corrections  \\
                &loop    & ($\frac{\alpha_{s}}{\pi}$ units)~~~~~~~~\\
  \hline
$\bar{\Pi}_{1}$  & $0.0084$   & $-0.0077$                                                                          \\ \hline
$\bar{\Pi}_{2}$  & $13.28$  & $-12.30$                                                                       \\ \hline
$\bar{\Pi}_{3}$  & $0.78$     & $-0.87$                                                                          \\ \hline
$\bar{\Pi}_{4}$  & $-2.25$    & $0.62$                                                                 \\ \hline
$\bar{\Pi}_{5}$  & $0.00$     & $0.20$                                                     \\ \hline
$\bar{\Pi}_{6}$  & $2.34$     & $-1.43$                                                                           \\ \hline
$\bar{\Pi}_{7}$  & $-0.097$   & $0.056$                                                                            \\ \hline
$\bar{\Pi}_{8}$  & $0.035$    & $0.41$                                                                         \\ \hline
$\bar{\Pi}_{9}$  & $0.623$    & $-0.87$                                                                           \\ \hline
$\bar{\Pi}_{10}$ & $1.72$     & $-1.61$                                                                             \\ \hline
$\bar{\Pi}_{11}$ & $0.696$   & $-1.04$                                                                            \\ \hline
$\bar{\Pi}_{12}$ & $0.165$    & $-0.16$                                                                            \\ \hline
Total            & $17.3$     & $-17.0$                                                                      \\ \hline
\end{tabular}
  \centerline{(b)}
\end{minipage}
\caption{(a) Numerical results for the massless quark-loop and the contributions from condensates. Table from \cite{Bijnens:2020xnl}. (b) Numerical results for the twelve different $\overline\Pi_i$ contributions from the quark-loop and the gluonic corrections in units of $10^{-11}$ and $10^{-11}\alpha_S/\pi$. Table from \cite{Bijnens:2021jqo}.}
  \label{tab:OPE}
\end{table}

\section{Perturbative corrections}

Since the nonperturbative corrections were small, the last place where large corrections might exist are the perturbative corrections, i.e. gluonic corrections to the massless quark-loop. A representative diagram is shown in Fig.~\ref{fig:quark-loop}(d). We used the method of master integrals and it turned out that all needed integrals are known analytically~\cite{Chavez:2012kn}. We used dimensional regularization for both IR and UV divergences, which all cancel. Results and more details can be found in~\cite{Bijnens:2021jqo}.

Numerical instabilities appear in the expressions near $\lambda= Q_1^4+Q_2^4+Q_3^4-2Q_1^2Q_2^2-2Q_2^2Q_3^2-2Q_3^2Q_1^2=0$ but one can perform the needed expansions analytically. Analytical results for the full expressions and all needed expansions are in the supplementary material of~\cite{Bijnens:2021jqo}.
Numerical results are shown in Tab.~\ref{tab:OPE}(b). Corrections are typically of order $-10\%$. The main uncertainty is which value of $\alpha_S$ to use, especially for lower $Q_{\text{min}}$. This is shown as the band in Fig.~\ref{fig:quark-loopnum}(b).

\section{Melnikov-Vainshtein limit}

A very strong constraint was derived in~\cite{Melnikov:2003xd} in the limit of two of the $q_i^2$ much larger than the third. Consequences for the theoretical amplitude are very strong because of the anomaly. Recent discussions of this can be found in \cite{Knecht:2020xyr,Ludtke:2020moa,Masjuan:2020jsf,Colangelo:2019lpu,Colangelo:2021nkr,Leutgeb:2019gbz,Cappiello:2019hwh}. In particular, it is known that the massless quark-loop reproduces this constraint~\cite{Bijnens:2021jqo,Colangelo:2019lpu} as argued earlier~\cite{Bijnens:2007pz}. There is also a short-distance gluonic corrections to this constraint, see~\cite{Ludtke:2020moa}. Our gluonic corrections reproduce this limit correctly.

\section{Conclusions}

We have done a study of the short-distance contributions for HLbL for the muon $g-2$. We have shown that this can be done properly in QCD and found that there are no unusually large corrections.

\acknowledgments
We thank Laetitia Laub for a fruitful and enjoyable collaboration.
This research is supported in part by the Albert Einstein Center for Fundamental Physics at Universit\"{a}t Bern (NHT), the Swedish Research Council grants contract numbers 2016-05996 and 2019-03779 (JB) and the Agence Nationale de la Recherche (ANR) under grant ANR-19-CE31-0012 (project MORA) (ARS).


\providecommand{\href}[2]{#2}\begingroup\raggedright\endgroup


\end{document}